\documentclass[preprint,amsmath,amssymb,aip,superscriptaddress,nofootinbib,showkeys]{revtex4-2}

\usepackage[english]{babel}
\usepackage[utf8]{inputenc}
\usepackage{xr-hyper}
\usepackage[colorlinks]{hyperref}
\usepackage[T1]{fontenc}
\usepackage[dvips]{graphicx}
\usepackage{amsmath}
\usepackage{amsfonts}
\usepackage{color}
\usepackage{multirow}
\usepackage[caption=false, justification=justified]{subfig}
\usepackage{url}
\usepackage{ulem}

\begin{document}

\title{Spin to charge conversion at Rashba-split SrTiO$_3$ interfaces from resonant tunneling.}

\author{D.~Q. To}
\email{duyquang.to@cnrs-thales.fr}
\affiliation{Unité Mixte de Physique, CNRS, Thales, Université Paris-Saclay, 1 Avenue Augustin Fresnel, 91767 Palaiseau, France}
\author{T.~H. Dang}
\affiliation{Unité Mixte de Physique, CNRS, Thales, Université Paris-Saclay, 1 Avenue Augustin Fresnel, 91767 Palaiseau, France}
 \author{L. Vila}
   \affiliation{Université Grenoble Alpes, CEA, CNRS, Spintec, F-38000 Grenoble, France}
  \author{J.~P. Attané}
  \affiliation{Université Grenoble Alpes, CEA, CNRS, Spintec, F-38000 Grenoble, France}
 \author{M. Bibes}
\affiliation{Unité Mixte de Physique, CNRS, Thales, Université Paris-Saclay, 1 Avenue Augustin Fresnel, 91767 Palaiseau, France}
\author{H. Jaffrès}
\email{henri.jaffres@cnrs-thales.fr}
\affiliation{Unité Mixte de Physique, CNRS, Thales, Université Paris-Saclay, 1 Avenue Augustin Fresnel, 91767 Palaiseau, France}

\date{\today}

\begin{abstract}
Spin-charge interconversion is a very active direction in spintronics. Yet, the complex behaviour of some of the most promising systems such as SrTiO$_3$ (STO) interfaces is not fully understood. Here, on the basis of a 6-band $\boldsymbol{k.p}$ method combined with spin-resolved scattering theory, we give a theoretical demonstration of transverse spin-charge interconversion physics in STO Rashba interfaces. Calculations involve injection of spin current from a ferromagnetic contact by resonant tunneling into the native Rashba-split resonant levels of the STO triangular quantum well. We compute an asymmetric tunneling electronic transmission yielding a transverse charge current flowing in plane, with a dependence with gate voltage in a very good agreement with existing experimental data.
\end{abstract}

\maketitle

\section{Introduction}
The interconversion between spin and charge currents mediated by Rashba interactions \citep{bychkov1984}, observed in a wide range of systems (Ag/Bi interfaces \citep{Sanchez2013}, SrTiO$_{3}$ two-dimensional electron gases (STO 2DEGs) \citep{Lesne2016,viret2016,Vaz2019,trier2019,vaz2020,bibes2021} CoFeB/MgO~\citep{viret2021}, Fe/Ge, $\alpha$-Sn \citep{Sanchez2013,Han2018}), offers huge technological opportunities from spin orbit torque and magnetic commutation to THz wave generation~\citep{Seifert2016, Jungfleisch2018, zhou2018,Akyildiz2020,dang2020}. In Rashba systems, a flow of charge results in an out-of equilibrium spin density arising from the uncompensated spin-texture at the Fermi surfaces, the so-called Rashba-Edelstein effect (REE)\citep{edelstein1990,raimondi2014,gorini2017}. Conversely, the inverse REE (IREE) allows for spin-charge conversion (SCC). Oxide interfaces based on SrTiO$_3$ (STO) have attracted a specific attention owing to their substantial Rashba field. Several works \citep{Lesne2016,viret2016,Vaz2019,bibes2020} have reported the observation of an enormous gate-tunable IREE in NiFe/LaAlO$_3$(LAO)/STO, NiFe/AlO$_x$(ALOx)/STO and related systems, with a SCC efficiency significantly larger than that of Ag/Bi(111)~\citep{Sanchez2013}. This makes STO 2DEGs promising candidates for the next-generation of high-speed and low-power spintronic devices such as magnetoelectric spin-orbit (MESO) transistors~\citep{Pai2018,Ohya2020,manipatruni2019,noel2020}.

However, the fundamentals and understanding SCC phenomena in 2DEGs oxide and heterostructures as provided in spin-pumping experiments is still in its infancy owing to the particular geometry measurements. In spin-pumping experiments performed on these systems~\cite{Lesne2016,Vaz2019}, a pure spin current is generated in a top magnetic contact of a tunnel devices, spin-current propagating in the confined STO layer constituting a quantum well (QW) before giving rise to spin-charge conversion \textit{via} assisted spin-orbit interactions (SOI). This makes largely not suitable the application of a conventional theory and modelling based on the linear Kubo's approach applied to the STO host matrix; unlike the issue of the reciprocal charge-to-spin conversion as required \textit{e.~g.} for the spin-torque problem.

In this article, we propose a modelling of SCC in the $\boldsymbol{k.p}$ framework giving rise to equivalent IREE phenomena in a STO electron gas (2DEG) confined in an oxide triangular quantum well (TQW) considering the specific structure of a NiFe/LAO/STO magnetic tunnel junction. We expose new theoretical insights on these phenomena from a quantum resonant tunneling point of view taking into account the symmetry breaking properties of Kramer's pair conjugates in STO QWs. We combine a $\boldsymbol{k.p}$ method and a scattering approach to describe spin-orbit assisted transport and SCC~\citep{Dang2015,Dang2018,To2019,Rozhansky2020} and demonstrate the occurrence of a lateral charge current with a specific gate dependence in very good agreement with experiments. We explain recent experimental data \citep{Lesne2016,Vaz2019} and go beyond existing tight-binding (TB) models \citep{zhong2013,khalsa2013,Vaz2019,Johansson2020} for tunnel structures, demonstrating the robustness of our method.

The paper is organized as follow. Section II focuses on the $\boldsymbol{k.p}$ model description and the calculations of the electronic band structure of STO in the presence of a Rashba surface potential. We compare our modelling to recent tight-binding (TB) results involving also the orbital structure. Section III is devoted to the calcualtion details of the resonant tunneling through Rashba states within STO quantum wells (QWs) together with their dependence on the gate voltage or electric field along the confinement direction (direction normal to interfaces). Section IV introduces the asymmetry of the tunneling transmission and discusses the associated spin-charge conversion responsible for the transverse charge current.

\section{STO Band structure with Rashba interactions.}

\subsection{Modelling.}

We start by describing the $\boldsymbol{k.p}$ electronic band structure of the STO host material.  TB~\citep{zhong2013,khalsa2013,shavanas2014,raimondi2016,Vaz2019,Johansson2020} as well as first principle theory~\citep{kong2021} have been extensively used to model the Rashba properties of STO 2DEGs. Nevertheless, although TB may correctly describe the Rashba energy splitting, it does not explicity deal with tunnelling structures. In addition, theories of SCC in oxide systems were limited, up to now, to in-plane charge current~\citep{raimondi2014,gorini2017,Johansson2020} whereas the spin-pumping technique, used in STO~\citep{Lesne2016,Vaz2019} as in semiconductor-based junctions~\citep{cerqueira2019}, implies a tunneling current normal to the layers. To this end, we follow the $\boldsymbol{k.p}$ approach of Heeringen~\textit{et al.}~\citep{Heeringen2013,Heeringen2017, Ho2019} and Ho~\citep{Ho2019}. One then considers the 6 following basis functions, \textit{namely} the two spin-degenerated light (\textit{le}) and heavy electrons (\textit{he}), as well as the two split-off (\textit{so}) components at the $\Gamma$ point ($\boldsymbol{k=0}$). Such characters (\textit{le}, \textit{he}, \textit{so}) are defined here along the quantification direction, $z$, normal to the layers. Note that they give rise to an opposite \textit{he} and \textit{le} character for the energy \textit{band} dispersions in the QW plane directions as found for semiconductors, wherein $X,Y$ and $Z$ refer to bonding \textit{p}-type symmetry orbitals in the latter case. The $\vert J,M \rangle$ states ($J$ the total angular momentum and $M$ its projection on the $z$ quantum axis) are then defined as:

\vspace{0.1in}

$\vert\frac{3}{2},\frac{3}{2}\rangle \equiv \vert le\uparrow\rangle=\frac{1}{\sqrt{2}} \vert (X+iY) \uparrow\rangle$

$\vert \frac{3}{2}, -\frac{1}{2}\rangle \equiv \vert he \uparrow\rangle = \frac{1}{\sqrt{6}}\vert (X-iY) \uparrow\rangle + \sqrt{\frac{2}{3}} \vert Z \downarrow\rangle$

$\vert \frac{3}{2}, \frac{1}{2}\rangle \equiv \vert he \downarrow\rangle = \frac{i}{\sqrt{6}} \vert (X+iY) \downarrow\rangle -i \sqrt{\frac{2}{3}} \vert Z \uparrow\rangle$

$\vert \frac{3}{2},-\frac{3}{2} \rangle \equiv \vert le \downarrow\rangle=\frac{i}{\sqrt{2}} \vert (X-iY) \downarrow\rangle$

$\vert \frac{1}{2},\frac{1}{2}\rangle \equiv \vert so \uparrow\rangle=\frac{1}{\sqrt{3}} \vert (X+iY) \downarrow\rangle + \frac{1}{\sqrt{3}} \vert Z \uparrow\rangle$

$\vert \frac{1}{2}, -\frac{1}{2} \rangle \equiv \vert so \downarrow\rangle=-\frac{i}{\sqrt{3}} \vert (X-iY) \uparrow\rangle + \frac{i}{\sqrt{3}} \vert Z \downarrow\rangle$
\vspace{0.1in}

\noindent where $\left\lbrace \vert X\rangle, \vert Y\rangle, \vert Z \rangle\right\rbrace \otimes \left\lbrace \uparrow,\downarrow \right\rbrace$ represents the tensorial product of the Ti 3\textit{d} orbitals $\left\lbrace d_{yz}, d_{zx},d_{xy}\right\rbrace \otimes\left\lbrace \uparrow,\downarrow \right\rbrace$ and $\uparrow, \downarrow$ the two spin states $\left\lbrace \uparrow,\downarrow \right\rbrace$. The $\boldsymbol{k.p}$ Hamiltonian is given \textit{via} a set of \textit{cubic parameters} namely $L,M,N$, the spin orbit splitting $\Delta_{SO}$ and a possible tetragonal distortion energy splitting $\Delta_{T}$~\cite{Heeringen2013} according to $\hat{\mathcal{H}}_{\boldsymbol{k.p}} = \hat{\mathcal{H}}_{cubic}\left(L,M,N\right) + \hat{\mathcal{H}}_{SO}\left(\Delta_{SO} \right) + \hat{\mathcal{H}}_{T}\left(\Delta_{T} \right) + \hat{\mathcal{H}}_{R}$ to give \textit{in fine}:

\begin{equation}
\small
H_{\boldsymbol{k.p}} = \begin{pmatrix}
p &b &-ia  &0 & \frac{1}{\sqrt{2}} a &-i\sqrt{2}b \\
b^{\dagger} &q &0 &ia &\sqrt{\frac{3}{2}}a^{\dagger} &c^{\dagger} \\
ia^{\dagger} &0 &q &b &c^{\dagger} &\sqrt{\frac{3}{2}}a \\
0 &-ia^{\dagger} &b^{\dagger} &p & -i\sqrt{2}b^{\dagger} &\frac{1}{\sqrt{2}} a^{\dagger} \\
\frac{1}{\sqrt{2}} a^{\dagger} &\sqrt{\frac{3}{2}}a &c &i\sqrt{2}b &r &0 \\
i\sqrt{2}b^{\dagger} &c &\sqrt{\frac{3}{2}}a^{\dagger} &\frac{1}{\sqrt{2}} a &0 &r
\end{pmatrix}
\label{k.pH}
\end{equation}

\vspace{0.2in}

\noindent with $a = \frac{1}{\sqrt{3}}N(k_{x}-ik_{y})k_{z}$, $b=\frac{1}{2\sqrt{3}}(L-M)(k_{x}^{2}-k_{y}^{2})$, $c =\frac{1}{3\sqrt{2}}i(L-M)(k_{x}^{2}+k_{y}^{2}-2k_{z}^{2})-\frac{\sqrt{2}}{3}i\Delta_{T}$, $p = \frac{1}{2}(L+M)(k_{x}^{2}+k_{y}^{2})+Mk_{z}^{2}$,
$q = \frac{1}{6}(L+5M)(k_{x}^{2}+k_{y}^{2}) + \frac{1}{3}(2L + M)k_{z}^{2} + \frac{2}{3}\Delta_{T}$,
$r = \frac{1}{3} (L+2M)(k_{x}^{2}+k_{y}^{2}+k_{z}^{2}) + \Delta_{SO} +\frac{1}{3}\Delta_{T}$. In the basis set of the order: $\left\lbrace \left\vert \frac{3}{2},\frac{3}{2} \right\rangle, \left\vert \frac{3}{2},-\frac{1}{2} \right\rangle , \left\vert \frac{1}{2},-\frac{1}{2} \right\rangle, \left\vert \frac{3}{2},-\frac{3}{2} \right\rangle , \left\vert \frac{3}{2},\frac{1}{2} \right\rangle, \left\vert \frac{1}{2},\frac{1}{2} \right\rangle \right\rbrace$, the $\boldsymbol{k.p}$ Hamiltonian (Eq.~[\ref{k.pH}]) can be re-written as:
\begin{eqnarray}
    H_{\boldsymbol{k.p}} = \begin{pmatrix} H_{\uparrow \uparrow} & H_{\uparrow \downarrow} \\
    H_{\downarrow \uparrow} & H_{\downarrow \downarrow}
    \end{pmatrix}
    ~\text{;} ~
        H_{\uparrow \downarrow} = Nk_{z}\begin{pmatrix} 0 &\frac{-i}{\sqrt{3}}k_{-} &\frac{1}{\sqrt{6}}k_{-} \\
    \frac{i}{\sqrt{3}}k_{-} &0 &\frac{1}{\sqrt{2}}k_{+} \\
    \frac{1}{\sqrt{6}}k_{-} &\frac{1}{\sqrt{2}}k_{+} &0
    \end{pmatrix}
\end{eqnarray}
with $k_{\pm}=k_{x}\pm ik_{y}$. We chose $L=0.6164$~eV\AA$^2$, $M=9.73$~eV\AA$^2$, $N=-1.615$~eV\AA$^2$, $\Delta_{SO}=28.5$~meV, $\Delta_{T}=2.1$~meV, obtained from a fitting procedure to Density Functional Theory (DFT)~\citep{Heeringen2013, Ho2019}. \textcolor{black}{One may possibly add an additional Bychkov-Rashba extra-term in the SrTiO$_3$ layer of the form:} $H_{R}=\alpha_{R}\left(\boldsymbol{\hat{\sigma}}\times \hat{\boldsymbol{p}}\right).\hat{z}$
as proposed in Ref.~\cite{petersen2000} where $\alpha_{R}$ is the Rashba parameter (the Rashba velocity $v_{R}$ is $v_{R}=\frac{\alpha_{R}}{\hbar}$), $\hat{z}$ is unit vector along $z$, $\hat{\boldsymbol{p}}$ is the momentum operator and $\hat{\boldsymbol{\sigma}}$ the Pauli matrices, one obtains the different Fermi surfaces for both the majority $\uparrow$ and minority spin $\downarrow$ channels.

\begin{figure}
\centering
    \includegraphics[width=.7\textwidth]{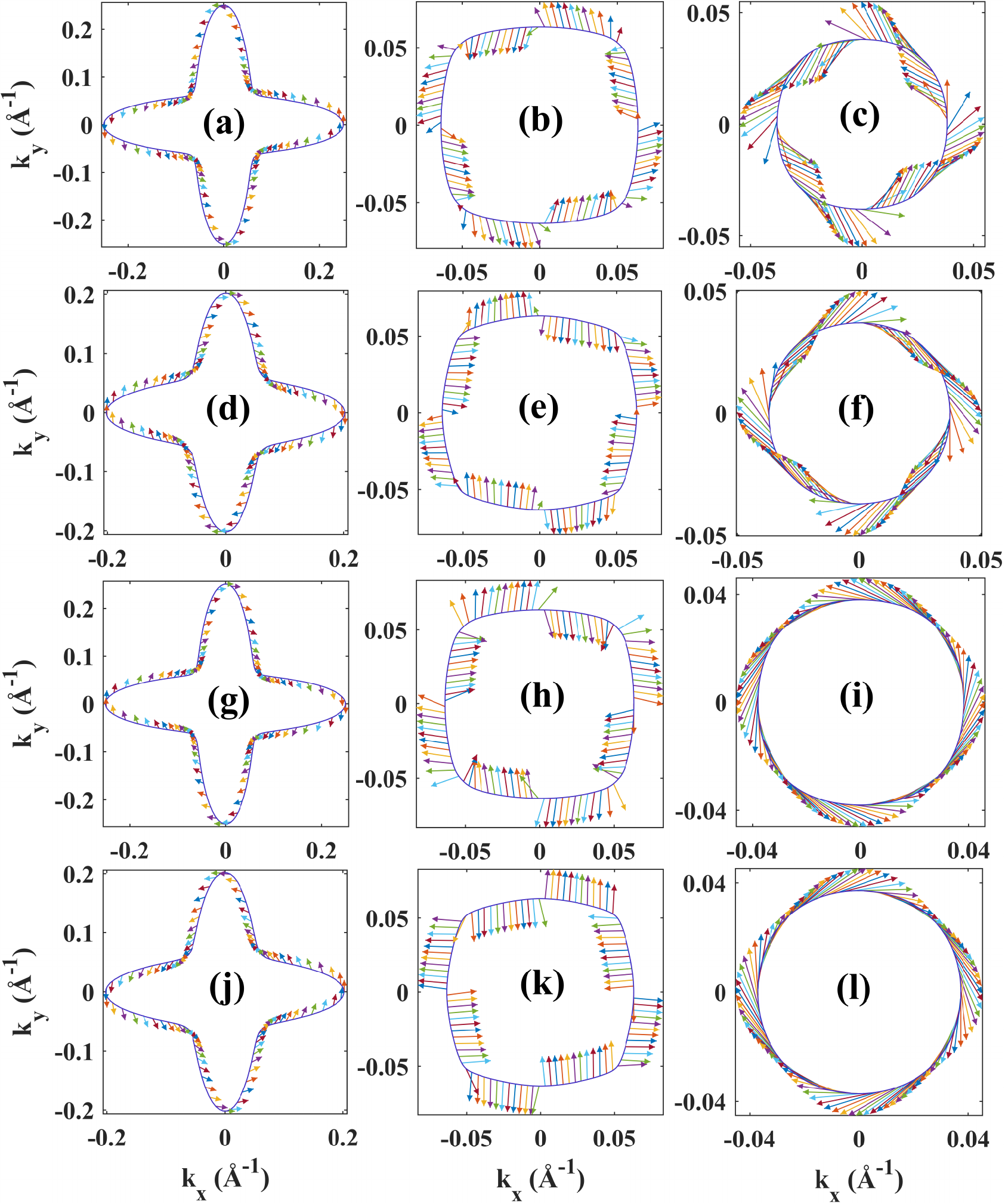}
 \caption{Fermi surface contours and (a-f): In-plane spin and (g-l) orbital  texture of le ($\uparrow,\downarrow$) (a,d,g,j), he ($\uparrow,\downarrow$) (b,e,h,k), and SO ($\uparrow,\downarrow$) (e,f,i,l) subbands of Rashba two dimensional electron gas at LAO/STO interface. The Fermi energy is chosen $\varepsilon=$0.04~eV above the bottom of CB. For each panels, the maximum size of the arrows corresponds to  $\frac{+\hbar}{2}$ for the spin and $\hbar$ for the orbit.}
  \label{FIG1}
\end{figure}

\subsection{STO Band structure involving Rashba potentials.}
\label{band_structure}

We now give a rapid description of the STO spin-resolved band structure in the presence of Rashba interactions where we fix now $\alpha_{R}=15$~meV.\AA. We compare our results to the ones of recent tight-binding (TB) treatment~\cite{Johansson2020}. Without magnetism, the resulting electronic bands of STO subject to an additional Rashba interaction are displayed on Fig.~\ref{FIG1}. Such additional Rashba term introduced is equivalent, in spirit, to the \textit{had-hoc} \textit{Rashba} hopping surface term proposed in TB~\cite{zhong2013,Johansson2020}.  The Fermi energy $\epsilon_F$ is chosen here to lie 0.04~eV above the bottom of the conduction band. The different bands and Fermi surfaces corresponds to respective $\uparrow$ (Fig.~\ref{FIG1}a, b and c) and $\downarrow$ spin channels (Fig.~\ref{FIG1}d, e and f), calculated from the above Hamiltonian. Figs.~\ref{FIG1}(a,d), Figs.~\ref{FIG1}(b,e) and Figs.~\ref{FIG1}(c,f) correspond respectively to \textit{he}$\uparrow,\downarrow$, \textit{le}$\uparrow,\downarrow$ and \textit{so}$\uparrow,\downarrow$ constituting the 6-band manifold. The heavy ($\textit{he}$), light ($\textit{le}$) and spin-orbit ($\textit{so}$) character of the bands are considered along the direction parallel to the electronic wavevector considering that the majority $\textit{he}$ eigenvectors slightly mix with $\textit{so}$ and vice-versa. In agreement with the cubic symmetry, one observes that the spin vectors remains orthogonal to the $k=(k_x,0)$ and $k=(0,k_y)$ directions as well as normal to the $k=(\pm k_x,\pm k_y)$ corresponding to each square diagonals. An in-plane helical spin texture thus emerge due to Rashba interactions with opposite winding for the spin channels; and such spin texture may be associated to Rashba-Edelstein (REE) as well as inverse related effects. Figs.~\ref{FIG1}g-l give the corresponding orbital texture associated to the 3\textit{d}-Ti orbitals showing the same type of helical structures. Note however that the correspondence between the spin and orbital textures for each band is not straightforward, between antiparallel (for the two first subbands) or parallel configurations (last subband). These results are in exact agreement with the TB calculations of Johansson \textit{et al.}~\cite{Johansson2020}.

\begin{figure}
         \centering
         \includegraphics[width=.65\textwidth]{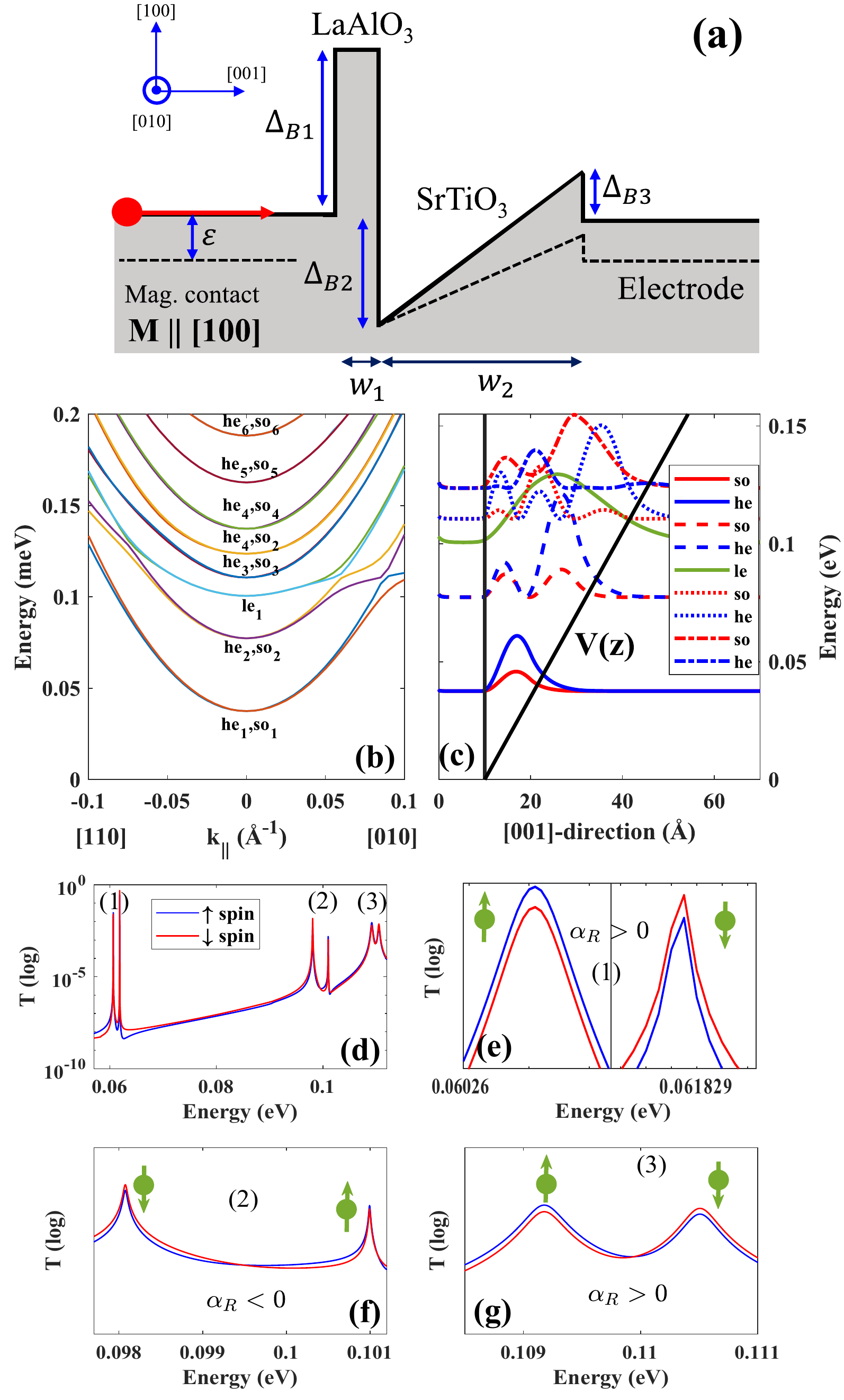}
\caption{(a) A ferromagnetic contact is deposited on top of a LAO/STO system forming a triangular well (TQW). (b) Energy dispersion in the 6~nm width STO TQW and \textit{he}1-\textit{SO}1, \textit{he}2-\textit{SO}2, \textit{le}1 subbands quantized levels. (c) Quantized wavefunction at the zone center ($k_\parallel=0$) for the first five levels with structural parameters given by $\Delta_{B1}$=2.5~eV, $\Delta_{B2}$=0.21~eV, $\Delta_{B3}$=0.05~eV, $w_{1}$=1~nm and $w_{2}$=6~nm. \textit{so}, \textit{he} and \textit{le} components are indicated in red, blue and green. (d) Transmission coefficient \textit{vs.} the incident kinetic energy $\epsilon$, of an electron tunneling through the structure shown in Fig.~\ref{FIG1}a with $k_{\parallel}=0.05$\AA$^{-1}$; (e-g) Resonant peaks $(1), (2), (3)$ depicted in Fig.~\ref{FIG1}d showing reverse spin texture of the levels.}
\label{FIG2}
\end{figure}

\section{Resonant tunnel theory with Rashba character.}

\subsection{Design of the tunneling structures.}

We now turn to the properties of a FM/I/STO tunnel device where 'FM' represent a 3\textit{d} ferromagnetic contact as represented \textit{e.g} by NiFe and 'I' the oxide barrier (typically LAO or AlO$x$). \textcolor{black}{We have considered the same type Hamiltonian for NiFe, $\hat{\mathcal{H}}_{\boldsymbol{k.p}} = \hat{\mathcal{H}}_{cubic} + \hat{\mathcal{H}}_{SO}\left(\Delta_{SO}\right) +\hat{\mathcal{H}}_{exc}$ without tetragonal distortion nor Rashba terms, but considering $\Delta_{SO}=70$~meV for 3\textit{d} elements and including an exchange splitting term $\hat{\mathcal{H}}_{exc}=-\Delta_{exc}~\hat{\sigma}.\hat{\mathbf{m}}$ with $\Delta_{exc}$=0.1~eV and a magnetization $\hat{\mathbf{m}}$ oriented along the $x=[100]$ in-plane direction. That way, as known from the spin-dependent tunneling process,} we are able to select one or two spin-channels in the tunneling process depending on the electron elastic energy.

\textcolor{black}{Unlike the aforementioned calculations of the STO band structure~\ref{band_structure} involving an explicit Rashba term ($\hat{\mathcal{H}}_R\neq 0$), the peculiarity of our tunneling approach will be the appearance of equivalent features from the native triangular potential with $\hat{\mathcal{H}}_R=0$, assumption that we will consider henceforth.} We will also simplify the self-consistent potential in STO considering a triangular form of the confined potential~\citep{Copie2009,Syro2011, Heeringen2013, Heeringen2017} $V(z) = -V_{0}+eF(z-z_{0})$ ($e$ is the charge), with $V_{0}=-\Delta_{B2}$ the potential at the STO interface $z=z_{0}$ and $F$ the electric field in the range $1-10$~meV/\AA~\cite{biscaras2014}. Moreover, we have considered here the same $\boldsymbol{k.p}$ cubic parameters for the barrier 'I' of height $\Delta_{B1}=2.5$~eV. The reference for the energy $\epsilon=0$ is taken 50~meV below the Fermi energy of the FM metal and matches with the bottom of the majority spin $\uparrow$ band. According to Ref.~\citep{baumberger2014} the band offset between the barrier and STO is $\Delta_{B2}=0.25$~eV at zero bias but may vary with \textit{a gate voltage} due to the charge transfer and its influence onto the self-consistent induced potential. For the following, the width of the TQW $w_2$ is kept fixed to 6~nm leading to an electric field $F$ of $F=4.2$~meV/\AA~ at $V_g=0$.

The structure under investigation is then sketched on Fig.~\ref{FIG2}a. The second barrier $\Delta_{B3}$ on the substrate side, and whose profile can be easily modified, is necessary for the formation of quantized states in the central STO layer and allows the collection of the current. Note however, that in the situation of spin-charge conversion induced by spin-pumping experiments as presently modelled, the collection of a longitudinal charge current in the STO substrate is not mandatory. Indeed, the spin-pumping current may be represented by a combined flux of majority spin $\uparrow$ to the right and equivalent minority spin $\downarrow$ flux to the left only causing, by transmission asymmetry effects, the transverse charge current in the plane of the quantum well we are searching for. This argument demonstrates the power of our present calculation method adapted to spin-pumping techniques.

\subsection{Quantum mechanics and electronic transmission.}

The wavefunction of the system that we have to solve is a solution of the Schr\"odinger equation:

\begin{equation}
\left[\hat{\mathcal{H}}_{\boldsymbol{k.p}}+\left(V\left(z\right)-\varepsilon \right)\hat{\mathcal{I}}\right] \psi\left(z\right)=0
\label{kpequa}
\end{equation}
that we have to solve without considering additional interface Rashba potential. At the vicinity of a given Rashba quantized level (or resonance $n$) at energy $\epsilon_n^\sigma$ in the TQW, the band-selected transmission coefficient $\mathcal{T}^{(n,\sigma)}\left(k_\parallel\right)$ ($n$ here is the band index and $\sigma$ denotes explicitly the spin) are of the form~\cite{doudin2006}:

\begin{equation}
\mathcal{T}^{(n,\sigma)} \left(k_\parallel\right)=\frac{4\Gamma_L^\sigma \Gamma_R}{\left[\epsilon-\epsilon_n^\sigma\left(k_\parallel\right)\right]^2+\left[\Gamma_L^\sigma+\Gamma_R\right]^2}
\label{resonant}
\end{equation}
where $\Gamma_L^\sigma\left(k_\parallel\right)$ and $\Gamma_R\left(k_\parallel\right)$ are respectively the spin-dependent energy broadening from the coupling to the FM and the unpolarized energy broadening towards the STO reservoir. $\Gamma_L^\sigma$ is sensitive to the spin eigenvalue of $n$ (respectively parallel or antiparallel to $x$) whereas $\Gamma_R$ is not. $\Gamma_L^\sigma$ will differ between the two split Rashba states for $k_\parallel$ along $\hat{z}\times \hat{m}$ in agreement with the symmetry rule required for the IEE process. $\Gamma_R$ generally larger $\Gamma_L$ depicts the electron lifetime $\tau_n=\frac{\hbar}{\Gamma_L}$ out of the 2DEG into STO. The voltage-integrated tunneling current on a given resonant level is $\tilde{\Gamma}=\int_0^{eV} \mathcal{T}^{(n,\sigma)} \left(k_\parallel\right) d(eV)=\frac{4\pi \Gamma_L^\sigma \Gamma_R}{\Gamma_L^\sigma +\gamma_R}$ with the results that $\tilde{\Gamma}=\frac{4\pi \Gamma_L^\sigma}{\Gamma_R}$ for $\Gamma_R\gg \Gamma_L^\sigma$ and $\tilde{\Gamma}=\frac{4\pi \Gamma_R}{\Gamma_L^\sigma}$ for $\Gamma_R\ll\Gamma_L^\sigma$. It results that the electronic transmission across a quantized Rashba state in STO and its hierarchy relative to their spin orientation may depends on the relative ratio $\frac{\Gamma_R}{\Gamma_L}$ so that, unlike conventional tunneling, a resonant transmission may invert the apparent sign of the Rashba interactions depending on the electron energy.

The calculation of the band-to-band selected transmission coefficients $\mathcal{T}_{k_\parallel}^{(n^\prime,n)}$  between incoming channel ($n$) and outgoing channel ($n^\prime$) (with $\mathcal{T}^{(n,\sigma)} \left(k_\parallel\right)=\Sigma_{n^\prime} \mathcal{T}_{k_\parallel}^{(n^\prime,n)}$)have been performed from the value of the band-selected transmission amplitudes $t^{(n,n^{'})}$ by using the multiple scattering Green's function formalism (evaluation of the S-scattering matrix computed from the scattering path operator). The discretization of the scattering region was performed into $N$ adjacent layers with $N=100$. The transmission amplitude at each interfaces between two consecutive regions have been found by using the standard matching conditions of continuity of the wavefunction and wavecurrent~\cite{Smith1990} without considering any additional surface Rashba terms.

\subsection{Results.}

Fig.~\ref{FIG2}b displays the energy dispersion in the STO TQW plane obtained for the different subbands (\textit{he}1, \textit{he2}, \textit{le}1, \textit{he}3, \textit{he}4,...) at $V_g=0$ along $[010]$ and $[110]$. One observes the appearance of an energy splitting for the first three subbands (\textit{he}1, \textit{he}2, \textit{le}1) relative to the two spins. This indicates a Rashba splitting without the use of any additional surface Rashba terms. Fig.~\ref{FIG2}c displays the quantized wavefunction for a strict normal incidence ($k_\parallel=0$) for the first five levels and showing a strong hybridization between the \textit{he} and \textit{so} components; the \textit{le} band remains pure. Indeed, a particular \textit{he-so} mixture leads to a pure $d_{xy}$ character of smaller mass near $\Gamma$ and able to minimize the quantization energy for the first two levels. The third level (\textit{le}1) retains a pure $d_{zx}$ character~\citep{Johansson2020}. Generally, at a finite wave vector $k_{\parallel}\neq 0$, each quantized state will be a mixture of \textit{he}, \textit{le} and \textit{so} subbands leading to a nonparabolic dispersion, especially at the vicinity of the anticrossing Lifshitz point. Here, the Rashba spin splitting reaches its largest value where the cubic Rashba spin-orbit term dominates the linear contribution close to $\Gamma$~\cite{Ho2019}.

Fig.~\ref{FIG2}(d-g) displays the resonant transmission spectra \textit{vs.} the elastic energy for $V_g=0$ at a slight oblique incidence $k_y=$~0.05\AA$^{-1}$. Fig.~\ref{FIG2}d shows the first three spin-split resonances respectively \textit{he}1 at 0.06~eV, \textit{he}2 at 0.1~eV and \textit{le}1 at 0.11~eV. The analysis reveals some trends about the sign of the effective Rashba interactions in STO. \textcolor{black}{Comparing the shape of the two peaks, either a given spin channel transmission peak (say the majority spin $\uparrow$ in blue) is broader/narrower than its opposite spin counterpart as on Fig.~\ref{FIG2}e (or Fig.~\ref{FIG2}f) or its amplitude is larger/smaller as on Fig.~\ref{FIG2}g.} From $(1)$ to $(3)$ on Fig.~\ref{FIG2}d-g, the Rashba coupling $\alpha_R^{(n)}=\frac{\Delta \epsilon_n}{|k_\parallel|}$ changes sign twice, as exhibited by the transmission for $\uparrow, \downarrow$ spins: it is respectively positive for both \textit{he}1 and \textit{le}1 and negative for \textit{he}2 with respective amplitudes $\alpha_R^{(1)}=$50~meV~\AA, $|\alpha_R^{(3)}|=$50~meV~\AA~ and $\alpha_R^{(2)}=$-100~meV~\AA~ as expected~\cite{baumberger2014}.

\begin{figure}
\centering
    \includegraphics[width=.75\textwidth]{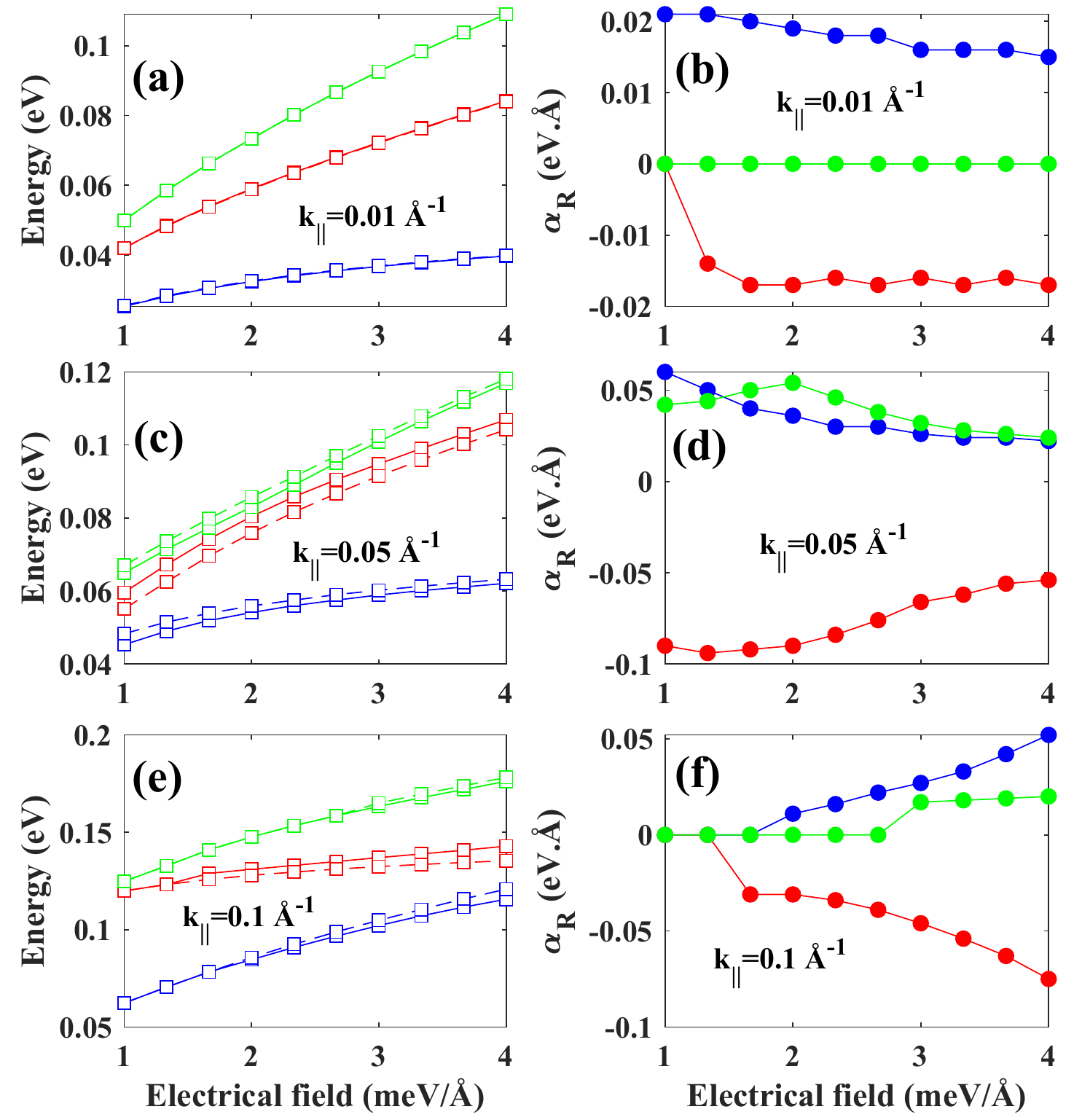}
 \caption{(a, c, e) The first three quantized levels of electron in an STO TQW \textit{vs.} the electric field calculated for different in plane wavevector $k_{\parallel}=0.01$~\AA$^{-1}$, 0.05~\AA$^{-1}$ and 0.1~\AA$^{-1}$. (b, d, f) effective Rashba parameters as a function of electric field obtained from splitting energy between spin up and spin down subbands according to $\alpha_{R}=\frac{\Delta \varepsilon}{\vert k_{\parallel}\vert}.$   }
  \label{FIG3}
\end{figure}

\vspace{0.1in}

\textit{Action of a gate-voltage:} We now focus on the action of a gate $V_g\neq0$ leading to a modulation of $F$ in STO. We model it by considering a change of the band offset $\Delta_{B2}$ between STO and the barrier with $F=\frac{\Delta_{B2}+\Delta_{B3}}{w_2}$. Fig.~\ref{FIG3} displays the resonance $\epsilon_n$ for the first three levels (\textit{he}1, \textit{he}2, \textit{le}1) and different $k_\parallel=0.01$\AA$^{-1}$ (Fig.~\ref{FIG3}a), $0.05$\AA$^{-1}$ (Fig.~\ref{FIG3}c), $0.1$\AA$^{-1}$ (Fig.~\ref{FIG3}e) as a function of $F$ in the range between 1 and 4~meV/\AA. For each $k_\parallel$ and for most of the resonances, $\epsilon_n$ admits an almost $\frac{1}{3}$ power law dependence on $F$ except for the second level ($n$=2) at $k_\parallel=0.1$\AA$^{-1}$~ because of the band degeneracy close to the Lifchitz point. Such dependence of $\epsilon_n$ is very close to what is theoretically expected from the relationship $\epsilon_n=\lambda_n \left(\frac{\hbar^2 e^2F^2}{2m^*}\right)^{\frac{1}{3}}$ with $\lambda_n=\left(\frac{3}{2}\pi(n-\frac{1}{4})\right)^{\frac{2}{3}}$ and $n$ the $n^{th}$ level for the given mass $m^*$ for \textit{he} and \textit{le} states. \textcolor{black}{We have gathered on table~\ref{table1} the values of the quantized energy extracted from our method and compared to the analytical theory given above with a good matching.} On Fig.~\ref{FIG3}(b,d,f), the effective Rashba parameter $\alpha_R^{(n)}=\frac{\Delta \epsilon_{(n)}}{k_\parallel}$ \textit{vs.} $F$ is plotted for different $k_\parallel$ (0.01, 0.05, 0.1~\AA$^{-1}$) and for the different subbands \textit{he}1 (blue), \textit{he}2 (red) and \textit{le}1 (green). At relative large $k_\parallel=0.1$~\AA$^{-1}$ $\alpha_R$ decreases largely on lowering $F$ down to 1~meV/\AA$^{-1}$ unlike at smaller $k_\parallel$. This has also been observed in the case of Schockley states~\cite{ishida2014}.

\begin{table}[h!]
\centering
\begin{tabular}{ |c||c|c||c|c|}
\hline
\multicolumn{5}{|c|}{Quantized energy in the TQW (eV)} \\
\hline
quantized level &
\multicolumn{2}{|c|}{$k_\parallel=0.01$~\AA$^{-1}$ ; F=1~meV/\AA} & \multicolumn{2}{|c|}{$k_\parallel=0.01$~\AA$^{-1}$ ; F=4~meV/\AA}\\
\hline
\hline
& this work & theory & this work & theory \\
\hline
\textit{he1} & 0.016 & 0.020 & 0.040 & 0.038 \\
\hline
\textit{he2} & 0.032 & 0.040 & 0.075 & 0.085 \\
\hline
\textit{le1} & 0.050 & 0.050 & 0.120 & 0.120 \\
\hline
\end{tabular}
\caption{\textcolor{black}{Quantized energy for \textit{he1, he2 and le1} levels calculated for two electric fields F=1~meV/\AA and F=4~meV/\AA and an in-plane wavevector $k_\parallel=0.01$\AA$^{-1}$. The left column corresponds to the values extracted by our method whereas the right column corresponds to the analytical theory for a TQW obtained for $m^*_{he}=10m_e$ and $m^*_{le}=0.7m_e$ for SrTiO$_3$.}}
\label{table1}
\end{table}

\section{Spin to charge conversion from resonant transmission asymmetry.}

\begin{figure}
\centering
    \includegraphics[width=.85\textwidth]{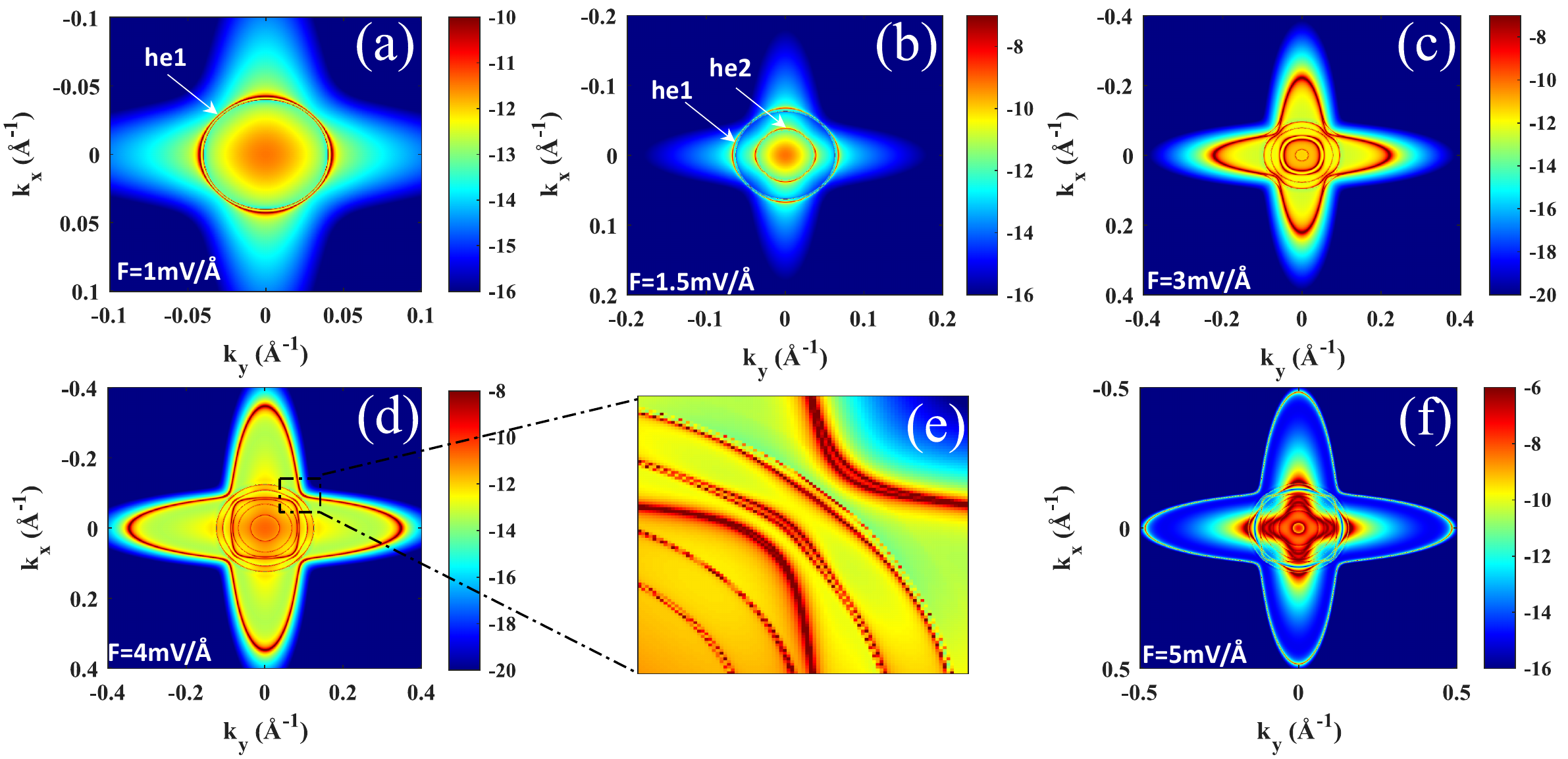}
 \caption{Transmission in color logarithm scale coefficient vs. the in-plane wavevector for electrons tunneling through LAO/STO TQW structure without magnetic contact. Such maps highlight the Fermi energy contour of quantized levels in the triangular quantum well of the cubic C4$_v$ symmetry ($600 \times 600$ grid k-points). Calculations have been performed for respective electric field $F=1, 1.5, 3, 4$ and 5 meV/\AA ~corresponding to respective panels a), b), c), e) (f is a zoom of e) and g). \textcolor{black}{(a) The case $F=1$~meV/\AA involves a single \textit{he1} spin-split Rashba resonance located at $\epsilon_R\approx 10$~meV from the bottom of the STO QW, and of Fermi wavevector $k_{F}^{he1}\simeq 0.043$~\AA$^{-1}$. (b) The case $F=1.5$~meV/\AA involves two Rashba-split resonances \textit{he1} ($\epsilon_R=20$~meV; $k_{F}^{he1}\simeq 0.065$~\AA$^{-1}$) and \textit{he2} ($\epsilon_R=35$~meV, $k_{F}^{he2}\simeq 0.038$~\AA$^{-1}$). For $F=3$~meV/\AA (c), tunneling involves 6 Rashba-split resonances, \textit{he1} ($\epsilon_R=35$~meV, $k_{F}^{he1}\simeq 0.23$~\AA$^{-1}$), \textit{he2} ($\epsilon_R=60$~meV, $k_{F}^{he2}\simeq 0.10$~\AA$^{-1}$), \textit{1e1} ($\epsilon_R=80$~meV, $k_{F}^{le1}\simeq 0.075$~\AA$^{-1}$), \textit{he3} ($\epsilon_R=80$~meV, $k_{F}^{he3}\simeq 0.055$~\AA$^{-1}$), \textit{he4-so} ($\epsilon_R=100$~meV, $k_{F}^{he4}\simeq 0.05$~\AA$^{-1}$) and \textit{he5-so} ($\epsilon_R=115$~meV, $k_{F}^{he5}\simeq 0.015$~\AA$^{-1}$).}}
  \label{FIG4}
  \end{figure}

We now turn to the calculation of the transmission coefficient in a FM/I/STO magnetic tunnel junctions. The calculation of the electronic transmission $\mathcal{T}$ was performed in the 2D reciprocal space as a function of the in-plane wavevector $k_\parallel=(k_x,k_y)$. Hereafter, we will consecutively switch-off and switch-on the exchange term $\Delta_{exc}$ in the left 'FM' magnetic contact and will compare the result of a non-magnetic (as played by Au or Al) and magnetic tunnel injector responsible for the appearance of a transverse charge current. We will consider different electric fields $F=1, 1.5, 3, 4$ and 5 meV/\AA; and in order to fit with the current values of $F$, the band offset band offset was respectively set to $\Delta_{B2}$ of 10, 40, 130, 190 and 250~meV with $\Delta_{B3}$ fixed at 50~meV.

The results obtained for a non-magnetic contact ($\Delta_{exc}=0$) are displayed Figs.~\ref{FIG4}(a-f) showing the resonance transmission at each Rashba quantized levels $\epsilon_R$ in agreement with our previous modelling. In each case, one may observe that the transmission map obey a perfect $C4_v$ cubic symmetry shape for electrons tunneling in the Brillouin zone as expected from the structure lattice structure (C4$v$ symmetry). \textcolor{black}{For $F=1$~meV/\AA ~(Fig.~\ref{FIG4}a), tunneling involves a single \textit{he1} spin-split Rashba resonance and giving rise to the same transmission at equivalent points on the cubic Fermi surface. For $F=1.5$~meV/\AA~(Fig.~\ref{FIG4}b), tunneling involves two Rashba resonance \textit{he1} and \textit{he2} Rashba-split bands. For $F=3$~meV/\AA (Fig.~\ref{FIG4}c), tunneling involves 6 different Rashba-split resonances, respectively \textit{he1}, \textit{he2}, \textit{1e1}, \textit{he3}, \textit{he4-so} and \textit{he5-so}.} When $F$ is further increased to $4-5$~meV/\AA, other resonances appear (Figs.~\ref{FIG4}d,f) but always fulfilling a cubic symmetry shape and thus providing an exact compensation of lateral charge current in the QW plane. Such symmetry in the transmission coefficient between two opposite incidences, $\pm k_y$, is expected from the time inversion operator and Kramer's pair conjugation in the STO TQW. Each spin $\uparrow$ state of in-plane incidence $+k_\parallel$ and energy $\epsilon=\epsilon_n (\uparrow,+k_\parallel)$ is conjugated to a corresponding spin $\downarrow$ state of in-plane incidence $-k_\parallel$  and same energy $\epsilon=\epsilon_n (\downarrow,-k_\parallel)=\epsilon_n (\uparrow,+k_\parallel)$. For a non magnetic system, an equivalent spin injection from the contact, it results an equal transmission \textit{vs.} $\pm k_\parallel$.

\begin{figure}
\centering
    \includegraphics[width=.75\textwidth]{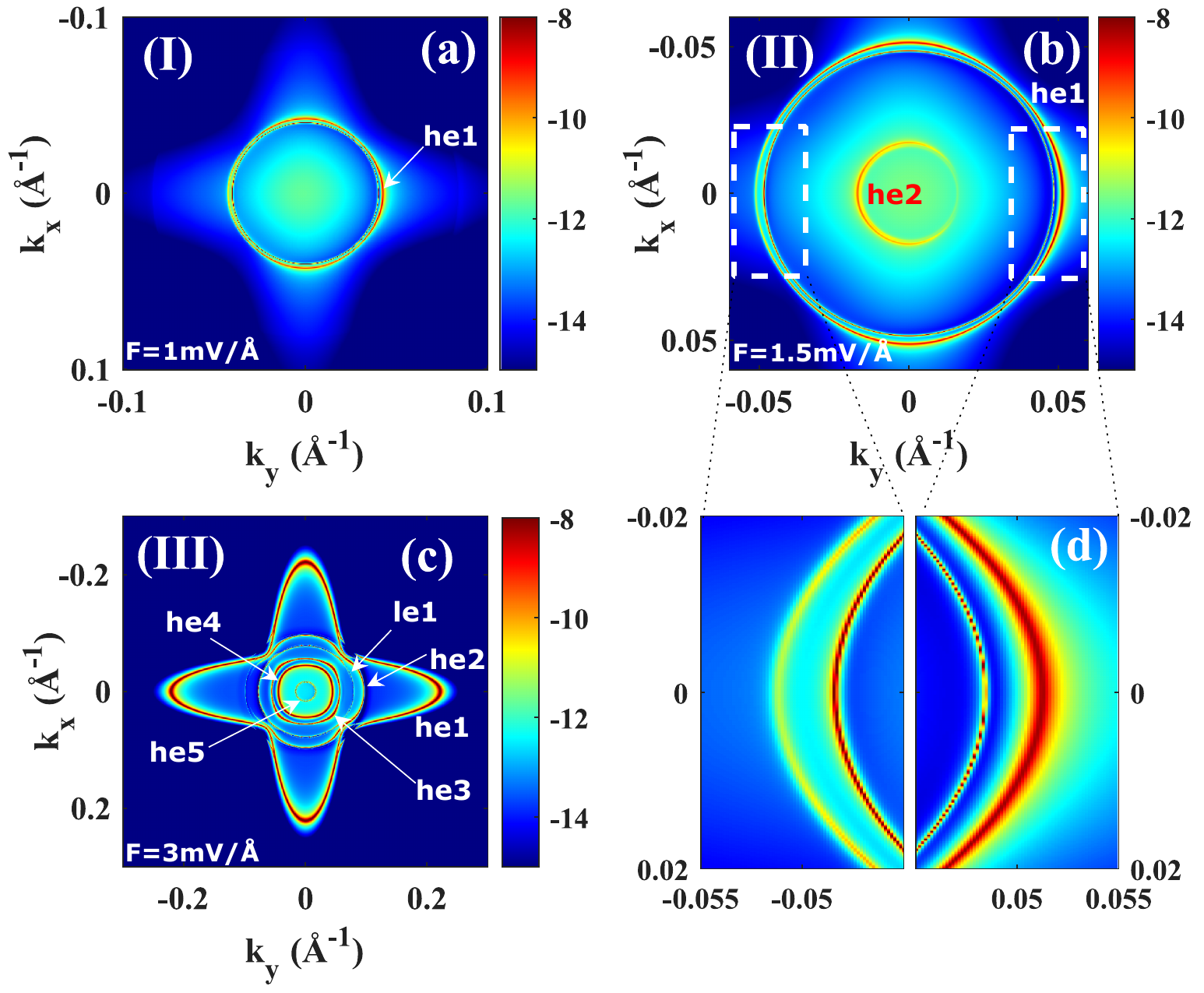}
 \caption{2D-map of the transmission coefficient (the color code represents the band-summed transmission in a log. scale) as a function of the in plane wavevector $k_{\parallel}=(k_{x},k_{y})$ ($600 \times 600$ k-grid points) for electron tunneling through LAO/STO TQW (a) with incident energy $\varepsilon=50$~meV (kinetic energy). The structural parameters are given in the text and matching with electric field at points $(I)$ \textcolor{black}{(1~meV/\AA)}, $(II)$ \textcolor{black}{(1.5~meV/\AA)} and $(III)$ \textcolor{black}{(3~meV/\AA)} in Fig.~\ref{FIG4}(b) for (a, b, c) respectively. (d) zoom of Fig.~\ref{FIG3}~(b) at $k_{y}=\pm 0.05$~\AA$^{-1}$ showing the largest Rashba spin splitting matching with the anticrossing point. \textcolor{black}{In each case, the character of the band, either \textit{he($n$)} or \textit{le($n$)}, is indicated emphasizing the sign of the corresponding effective Rashba splitting.}}
  \label{FIG5}
\end{figure}

We now switch on the exchange parameter $\Delta_{exc}=0.1$~eV. The result is different due to the lift of the spin degeneracy of the Kramers's pair within the TQW in the electronic transmission~\cite{fasolino1992}. We calculate the corresponding transmission map as the same time that the transverse charge current for $F=1, 1.5, 3$~meV/\AA. A magnetic contact allows an asymmetry of the transmission as a function of the carrier incidence with respect to the reflection plane defined by the magnetization $\mathbf{M}$ and the surface normal. Due to the axial character of $\mathbf{M}$, two electrons with exact opposite in-plane wave vector $\pm k_{\parallel}$ may have different transmission amplitude~\cite{Dang2015,Dang2018,To2019,Rozhansky2020} unlike the case of non-magnetic systems. A charge current flowing in-plane is then expected to occur along the $y=[010]$ direction. The results are displayed on Fig.~\ref{FIG5} respectively for $F=1$~meV/\AA~(Fig.~\ref{FIG5}a:~point $I$), $F=1.5$~meV/\AA~ (Fig.~\ref{FIG5}b:~point $II$) and $F=3$~meV/\AA~ (Fig.~\ref{FIG5}c:~point $III$). The potential profiles \textit{vs.} $F$ still match with an elastic resonant tunneling injection into $I$) the first \textit{he1} Rashba split bands ($\alpha_R>0$ with $k_{F}^{he1}\simeq 0.043$~\AA$^{-1}$),  into $II$) the two first \textit{he}1 ($\alpha_R>0$) and \textit{he2} ($\alpha_R<0$) bands ($k_{F}^{he2}\simeq0.015$~\AA$^{-1}$) with opposite sign of Rashba and into $III$) the \textit{he1} ($k_{F}^{he1}\simeq0.225$~\AA$^{-1}$), \textit{he2} (\textcolor{black}{$k_{F}^{he2}\simeq0.095$}~\AA$^{-1}$), \textit{le1} (\textcolor{black}{$k_{F}^{le1}\simeq0.055$}~\AA$^{-1}$), \textit{he-so3} and \textit{he-so4} bands.

\begin{figure}
\centering
    \includegraphics[width=.75\textwidth]{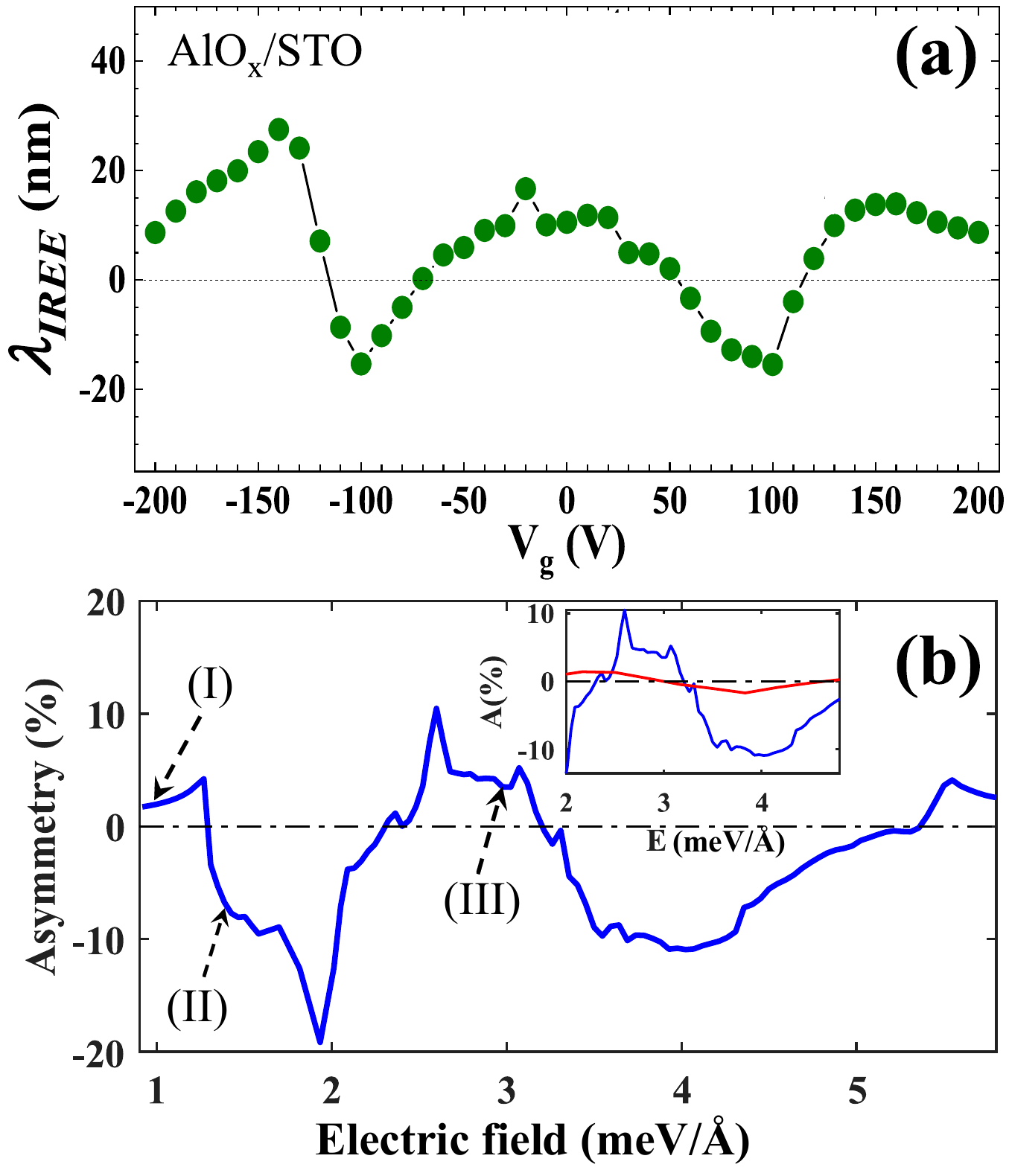}
\caption{(a) Spin-charge conversion efficiency $\lambda_{IREE}$ \textit{vs.} $V_{g}$ for ALO/STO structure (at T=15K) taken from Ref.~\citep{Vaz2019}. (b) The asymmetry \textit{vs.} the electric field $F$ calculated with $\varepsilon=50$~meV. The inset figure presents the oscillation of asymmetry for two case: $\Delta_{B3}=50$~meV and $\Delta_{B3}=+5$~meV for blue and red curves, respectively. These results were performed with $600 \times 600$ grid points of in-plane wave vector $k_{x} \times k_{y}$.}
 \label{FIG6}
\end{figure}

One observes now that, involving non-zero magnetism in the injector ($\Delta_{exc}=0.1$~eV), the transmission does not obey a perfect cubic symmetry shape but differs for $k_\parallel=\pm k_y$ in the direction $\hat{z}\times \hat{m}$ as expected (see \textit{e.g} Fig.~\ref{FIG5}d). The transmission along $\pm k_x$ remains symmetric for fixed $k_y$. Concomitantly to a tunneling spin-current along $z$, a different transmission along $+k_y$ and $-k_y$ has to be associated with a transverse charge flow along $y$ describing an IREE. In the case $I$ the transmission is larger along $+k_y$ for the first \textit{he1} band which defines $\alpha_R^{(1)}>0$. For the case $II$, the inner Rashba band with smaller $k_F$ gives an opposite sign to the transmission asymmetry and this should be linked to $\alpha_R^{(2)}<0$ as previously mentioned. Note that in this tunneling geometry, the inner Rashba band gives an overall larger conduction owing to the reduced incidence (smaller $k_F$) and then to an overall negative IRE effect. In the case $III$, the first Rashba band gives a standard Fermi 'cigar' shape to the resonant transmission however assigned to a very small selected $\pm k_y$ asymmetry. In the case $III$, additional Rashba split band at higher energy (\textit{le}1, \textit{he}3-\textit{so}, \textit{he}4-\textit{so}) of a positive Rashba signature yields an overall positive IRE ($J_{+k_y}>J_{-k_y}$). These conclusions are compatible with the description of Fig.~\ref{FIG3} in particular in terms of the Rashba coupling sign. Such transmission difference \textit{vs.} $\pm k_y$ is characterized by the asymmetry parameter $\mathcal{A}$ as:
\begin{equation}
    \mathcal{A}\left(\varepsilon\right) = \frac{\sum\limits_n\sum\limits_{k_{x}}\sum\limits_{k_{y}>0}\left[T\left(\varepsilon,k_{x},+k_{y}\right)-T\left(\varepsilon,k_{x},-k_{y}\right)\right]}{\sum\limits_n\sum\limits_{k_{x}}\sum\limits_{k_{y}>0}\left[T\left(\varepsilon,k_{x},+k_{y}\right)+T\left(\varepsilon,k_{x},-k_{y}\right)\right]}
\end{equation}

Chirality effects of the same origin have been discussed in terms of tunneling anomalous Hall effects or TAHE in semiconductors~\cite{Fabian2015,Rozhansky2020} or in superconducting materials~\cite{Fabian2019}. In this paper, we generalize these phenomena to resonant tunneling in oxide based systems. A non zero $\mathcal{A}$ gives rise to a transverse charge current $\mathcal{J}_{c}$ flowing in the QW plane with $\mathcal{J}_{c} \approx \mathcal{A}w_2\sin\left(\theta_{k_\parallel}\right)\mathcal{J}_{s}$
where $w_2$ is the QW width, $\mathcal{J}_{s}$ the spin current along $\hat{z}$, as computed by our $\boldsymbol{k.p}$ method and $\theta_{k_\parallel}\lesssim 1$ the spin-current injection angle from $\hat{z}$. Note however that $\mathcal{A}\propto \mathcal{P}$ is proportional to the spin-current polarization injected from the contact, and then proportional to the spin-density (or spin accumulation $\mathcal{P}$) injected in the STO TQW. This gives the connection we are searching for between the transverse charge current and the spin-polarization in the STO, equivalent to IREE. In the present case, calculations give an an out-of-equilibrium spin-polarization $\mathcal{P}\approx 0.6$in the STO QW textcolor{red}{(\textit{not shown})} with the chosen parameters. Whereas the charge current $\mathcal{J}_c$ remains constant within the tunneling structure, $\mathcal{J}_s$, of vectorial nature, may vary in the STO TQW owing to the Rashba field inducing a local spin-precession (the Rashba interaction does not commute with the spin operator). Fig.~\ref{FIG6}b displays $\mathcal{A}$ \textit{vs.} $F$ for an electron beam incoming with incident energy $\varepsilon=0$ (kinetic energy = 50~meV) from the band bottom. Upon the increase of $F$ above 1~meV/\AA~, $\mathcal{A}$ starts with small positive value of about $+2\%$ corresponding to a weak Rashba splitting of the first quantized level ($I$ in Fig.~\ref{FIG5}a). Then, $\mathcal{A}$ decreases down to a large negative due to an opposite spin textures of the second level for $F>1.2$~meV/\AA. In this region, the asymmetry reaches a maximum absolute value of $-20\%$ ($II$) at the largest Rashba splitting at the vicinity of the Lifshitz point where the current spin-polarization within the TQW approaches unity. Then, when $F$ is further increased, $\mathcal{A}$ changes its sign to become positive again ($III$) when the third and upper levels get involved in the tunneling. This evolution of $\mathcal{A}$ perfectly reproduces in shape the trend of spin to charge conversion length, $\lambda_{IREE}$, \textit{vs.} bias $V_g$ reported in experiments of Vaz \textit{et al.} \citep{Vaz2019} for ALO$x$/STO (Fig.~\ref{FIG6}a). Fig.~\ref{FIG6}b gives the variation of $\mathcal{A}$ when the right tunnel barrier is reduced to a minimum value $\Delta_{B3}=5$~meV making the energy broadening to the right reservoir $\Gamma_R$ very large and the outward tunneling time $\tau_R=\frac{\hbar}{\Gamma_R}$ very short. \textcolor{black}{The strong reduction of $\mathcal{A}$ observed should be assigned to the reduction of the particle lifetime $\tau_R$ limiting thus the efficiency of the SCC in agreement with the standard diffusive model of IREE~\cite{Sanchez2013} in the regime of current-in-plane (CIP) injection.}

\section{Conclusions.}

\textit{Via} a 6 band $\boldsymbol{k.p}$ spin-dependent resonant tunneling model, we have demonstrated the gate dependence of the spin-charge interconversion mechanism in an insulator/STO triangular quantum well for perpendicular spin-current flow, in very good agreement with spin pumping experimental data. This is demonstrated without the need to add a supplementary interface Rashba potential. This behavior of tunneling asymmetry in LAO/STO system also demonstrates the universal phenomenon of chirality-driven skew scattering~\cite{Rozhansky2020} involving strong electric fields or potential gradients and subsequent spin-orbit interactions. A deeper analysis will be required within this $\boldsymbol{k.p}$ formalism to go beyond for further conclusions in fundamentals of SCC with STO. In particular, exploring the role of the orbital Edelstein effect, predicted to be strong in STO 2DEGs ~\cite{Johansson2020}, and the influence of ferroelectricity~\cite{noel2020}, appear as interesting directions for the future.

\vspace{0.2in}

\begin{acknowledgments}
This work received support from the ERC Advanced Grant n° 833973 'Fresco'. We acknowledge financial support from the Horizon 2020 Framework Programme of the European Commission under FET-Open Grant No. 863155 (s-Nebula). We acknowledge the ANR program 'ORION' through grant number ANR-20-CE30-0022-01.
\end{acknowledgments}


\end{document}